\documentclass[prb,twocolumn,showpacs,preprintnumbers,amsmath,amssymb,superscriptaddress]{revtex4}

\usepackage{graphicx}
\usepackage{dcolumn}
\usepackage{amsmath}

\begin{document}
\title{Structure of self-organized Fe clusters grown on Au(111) analyzed by Grazing Incidence X-Ray Diffraction}

\bibliographystyle{apsrev}
\author{H. Bulou}
\email{bulou@ipcms.u-strasbg.fr}
\author{F. Scheurer}
\affiliation{Institut de Physique et Chimie des Mat\'{e}riaux de
Strasbourg, UMR 7504 CNRS-Universit\'{e} Louis Pasteur, 23 rue du
Loess, BP 43, F-67034 Strasbourg Cedex 2 (France) }

\author{P. Ohresser}
\affiliation{Laboratoire pour l'Utilisation du Rayonnement
Electromagn\'{e}tique, UMR 130 CNRS-Universit\'{e} Paris-Sud, Bat
209 D, F-91898 Orsay Cedex (France)}

\author{A. Barbier}
\affiliation{CEA-Saclay DSM/DRECAM/SPCSI, F-91191 Gif-Sur-Yvette
(France)}

\author{S. Stanescu}
\affiliation{European Synchrotron Radiation Facility, BP 220
F-38043 Grenoble Cedex (France)}
\author{C. Quir\'os}
\affiliation{European Synchrotron Radiation Facility, BP 220
F-38043 Grenoble Cedex (France)}

\preprint{DRAFT - Version 3.5}

\date{\today}

\begin{abstract}
We report a detailed investigation of the first stages of the growth of self-organized Fe
clusters on the reconstructed Au(111)
surface by grazing incidence X-ray diffraction. Below one
monolayer coverage, the Fe clusters are in "local epitaxy" whereas
the subsequent layers adopt first a strained fcc lattice and then
a partly relaxed bcc(110) phase in a Kurdjumov-Sachs epitaxial
relationship. The structural evolution is discussed in relation
with the magnetic properties of the Fe clusters.
\end{abstract}


\maketitle
\section{Introduction}

Nanometer-sized objects like metallic clusters or
wires grown by controlled self-organization
processes on surfaces exhibiting a defined strain-field
(\emph{e.g.} reconstructed surfaces and patterned surfaces) are
currently under extensive study. Because of the reduced
dimensionality and the inflated importance of surface phenomena
these new objects exhibit original properties. In particular, they
are models for investigating the magnetic properties of low
dimensionality systems.\cite{Gambardella02, Ohresser01, Shen03}
As a matter of fact, the magnetic properties are closely related
to the intimate crystalline structure. Very small objects can
present new and/or highly strained crystallographic phases with
respect to the bulk equilibrium ones. These phases may in turn
lead to peculiar spin phases or magnetic anisotropies.
Investigating the structure of such small objects is therefore
important but very difficult because of the very small amount of
deposited material. Recently, the magnetic properties of
self-organized Fe deposits on the reconstructed Au(111) surface
were investigated by X-ray Magnetic Circular Dichroism (XMCD).
Three different spin-phases were identified, as a function of
coverage.\cite{Ohresser01} To fully understand these magnetic
properties it is mandatory to precisely determine the crystalline
structure of the Fe deposits. For such small clusters grown on a
surface with a large lattice mismatch, one may expect strong
differences compared to the bulk structure. 

In this paper, the
crystalline structure and strain relaxation of Fe deposits on
Au(111) is investigated by Grazing Incidence X-Ray Diffraction
(GIXD) as a function of Fe coverage. 
In a first part, the characteristics of the reconstructed Au(111)
surface are given, and the reciprocal lattice is analyzed by GIXD. In a second part 
the structure of Fe layers by GIXD as a function of coverage is studied.
The structural evolution of these Fe clusters is then discussed with
respect to the structural and morphological properties of the
reconstructed Au(111) surface.
The analyzed Fe thicknesses
range from isolated and self-organized clusters to several
monolayers (ML). It is shown that Fe clusters first grow in close
registry on the highly inhomogeneous reconstructed Au(111)
surface, then after the coalescence, a relaxation starts upon the
growth of the second layer. Above 3 ML, the film
undergoes a phase transition from a fcc(111) phase to a bcc(110) phase.

\section{Experimental}

The experiments were performed in ultra-high vacuum conditions
on the Surface
Diffraction Beamline (ID3) at the ESRF on a z-axis
diffractometer.\cite{FCO95,ESRFWWW} The single crystalline Au
substrate was of (111) orientation within $\pm$0.1$^{\circ}$. It
was prepared in situ by Ar$^{ + }$ sputtering and annealing
cycles, up to 1000 K. The Fe was evaporated from a high
purity-rod, heated by electron bombardment in an evaporation cell
equipped with a flux monitor. During the evaporation the pressure
was in the low $10^{-10}$ mbar range. The growth was made with the
substrate held at room temperature. After each analyzed coverage,
the sample was cleaned and the Fe deposition repeated. The
thickness of the deposit was controlled by fitting the specular
reflectivity Kiessig fringes. The surface and Fe deposits
cleanliness were controlled by Auger Electron Spectroscopy. The
(111) single crystalline surface was described by the classical
triangular unit cell.\cite{SMZ91,Grubel} defined by the surface
in-plane basis vectors $\vec{a}_1$, $\vec{a}_2$, making a
120$^\circ$ angle ($a_1$=$a_2$=$a_0/\sqrt{2}$, where $a_0$= 2.88 {\AA} is the
fcc bulk parameter of gold) and $\vec{a}_3$, perpendicular to the surface
($a_3=\sqrt{3}a_0$). In this way the reciprocal space indices $H$
and $K$ describe the in-surface plane momentum transfer, and $L$
the perpendicular to the surface momentum transfer. The photon
energy was set to 17.176 keV. The angular resolution for in-plane
scans was 0.03 mdeg and the incidence angle was tuned close to the
value for total external reflection of the X-rays ($\sim$0.3
${^\circ}$ at 17.176 kev).

\section{Au(111) surface reconstruction}

\begin{figure}[htbp]
\includegraphics[width=9cm,angle=0]{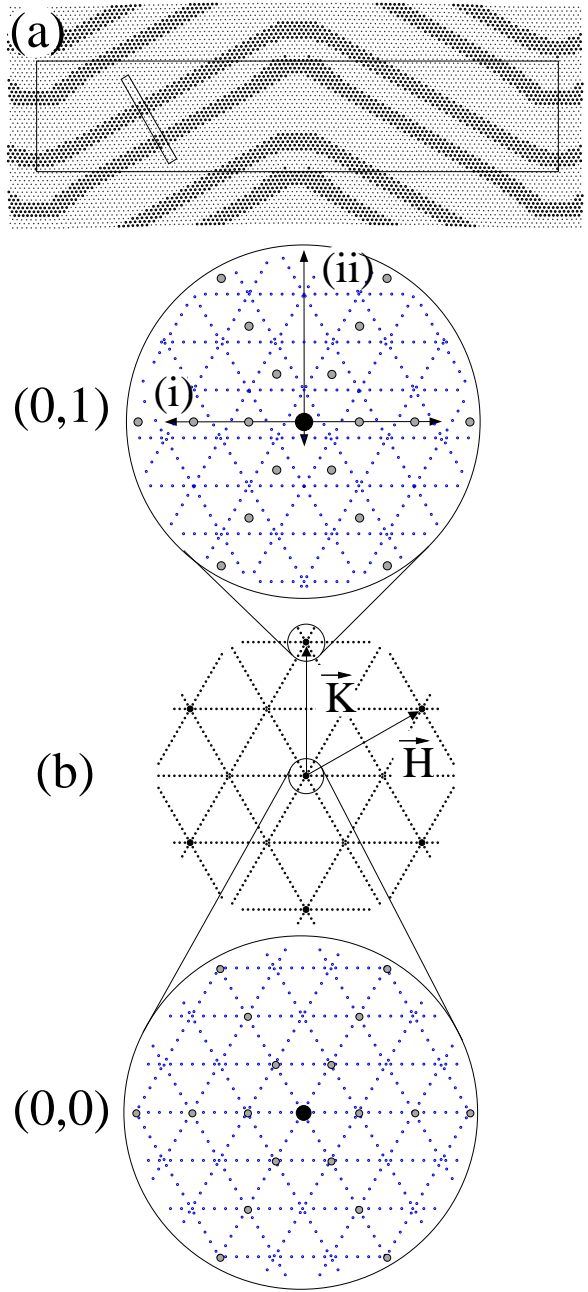}
\caption{(a) Reconstructed Au(111) surface as simulated by Molecular Dynamics.\cite{BGO02}
The $22\times\sqrt{3}$ reconstruction cell is represented (small rectangle), as well 
as the rectangular herringbone unit cell (large rectangle).  In this simulation the large periodicity of the kink array 
is L=32 nm (see text).  
(b) Reciprocal surface lattice of reconstructed Au(111) (middle) with enlargements around 
the (0,0) (bottom) and (0,1) (top) reflections. The intersection of the crystal truncation rods with the surface plane are represented by the dark 
disks, the $22\times \sqrt{3}$ reconstruction by smaller grey disks and the reflections due to the kink 
lattice by dots (for sake of clarity, the contribution of the kinks are only represented in the enlarged regions). 
The different types of GIXD scans that have been recorded in the present study are indicated by arrows as(i) and (ii).}
\label{fig:RRAu111}
\end{figure}

In order to understand the modifications in the X-ray diffraction 
scans induced by the evaporation of very small amounts of Fe it is
important to first precisely describe the Au(111) reciprocal
lattice. 
The strong relativistic effects experienced by the
electrons in gold produce a large mismatch between the bulk
equilibrium interatomic distance and the surface one.\cite{relat,BGO02}
Moreover, since the chemical nature of the surface and bulk atoms is the same,
the interactions between two surface atoms and between a surface and a bulk atom have
approximately the same magnitude. 
Hence, for the surface atoms, a competition results between the trend to get closer, or to adopt
the bulk equilibrium distance.
 In the case of the Au(111) surface, the most favorable energetic situation consists of a
surface split into domains of two different types: domains where
the interatomic distance is the one of an ideal gold surface (hcp
stacking) and domains with the bulk interatomic distance (fcc stacking). 
The two types of domains are separated by
discommensuration lines, several atomic distances wide, which make
the junction between the hcp and the fcc domains. 
The interatomic distances between surface atoms are very inhomogeneous since they vary from 2.65 \AA\  to 2.86 \AA.\cite{BGO02} 
The periodic succession of parallel fcc and hcp stripe
domains along the denser $<\bar{1}01>$
atomic direction forms an uniaxial $22\times\sqrt{3}$
reconstruction. 
The density increase along $<\bar{1}01>$ rows (by introducing one additional Au atom every 22 bulk atoms) produces stress in the other symmetry 
equivalent directions (Au(111) belongs to the \emph{F}m3m space group). 
Hence the stripe domain reconstruction is unstable for large areas\cite{alerhand,marchenko}  and the best compromise is the formation of three 
types of stripe domain reconstructions, each of them associated with one of the three equivalent $<\bar{1}01>$ directions. 
The intersection between the discommensuration lines of the different stripe domains induces the formation of kinks. The kinks are themselves 
ordered, and this leads to a structure in which two of three possible rotational equivalent domains of the stripe domain structure alternate periodically 
across the surface, forming the well-known zigzag pattern.\cite{Barth,SMZ91,Vanderbilt} 

In summary, the herringbone-reconstructed Au(111) surface can be understood as the
superposition of three different lattices (figure \ref{fig:RRAu111}a):
\begin{enumerate}
    \item The fcc bulk lattice with lattice parameter $a_{0}$.
    \item The surface reconstruction lattice,
used to be called $22\times \sqrt{3}$, which is the consequence
of the density increase along the $<\bar{1}01>$ rows. In a
$<\bar{1}01>$ direction there is a 22a$_{0}$ super-periodicity,
whereas in the perpendicular $<1\bar{2}1>$ direction, the
super-periodicity is $\sqrt{3}a_{0 }$.
    \item The rectangular ($l \times L$) kink super-lattice with
$l$=7.2 nm (fixed precisely by the 22a$_{0}$ reconstruction
periodicity), and $L$ varying typically between 15 nm and 50 nm.
The length $L$ experiences very large fluctuations
from one sample to another, depending on the
preparation conditions and the density of defects in the crystal.\cite{Fruchart03}
\end{enumerate}

The resulting Au(111) reciprocal space is made of the
superposition of the reciprocal lattices of the three different
lattices and the symmetry-equivalent domains (figure
\ref{fig:RRAu111}b). The large disks are the intersections with the
(H,K) plane of the crystal truncation rods (CTR),\cite{Robinson}
the small grey disks represent the lattice of the $22\times
\sqrt{3}$ reconstruction, and the dots represent the theoretical
lattice formed by the regular arrangement of the kinks.

The Au(111) surface reconstruction has already been investigated
in detail by GIXD with a high angular resolution by Sandy
\emph{et al.}.\cite{SMZ91} Note however that the reciprocal space
represented in Fig. 6 of ref. \onlinecite{SMZ91} is in fact only valid
around the specular reflection: since the rectangular lattice is
incommensurate, one does not have exactly the same positions of
the kink reflections around a (0 1) reflection (see figure
\ref{fig:RRAu111}b). Indeed, the representation of our figure
\ref{fig:RRAu111}b corresponds much better to the experimental scans
of Sandy \emph{et al.} (figure 8 of ref. \onlinecite{SMZ91}).

\begin{figure}[htbp]
\includegraphics[width=8cm,angle=0,bb=60 200 490 640]{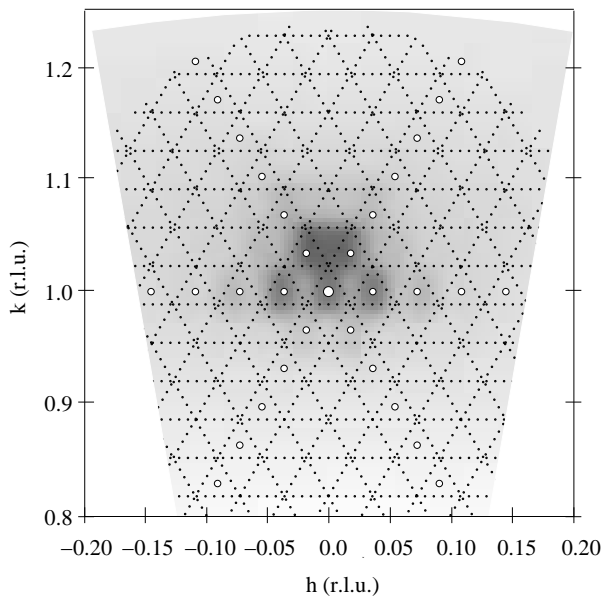}
\caption{\label{fig:RRAu111Map}Experimental map around the (H K L) = (0 1 0.12)
region of the reciprocal space for clean gold. (r.l.u. stands for reciprocal
lattice units). The map is obtained by interpolating rocking-scans recorded as a function of K-steps. The calculated reciprocal lattice of the gold 
reconstruction is superimposed. The dots stand for kinks and the $22\times\sqrt3$ is represented by small open circles. The black central dot is the 
intersection of the (0 1 L) crystal truncature rod with the surface plane}
\end{figure}

\begin{figure}[htbp]
\includegraphics[width=8cm,bb=50 30 565 800]{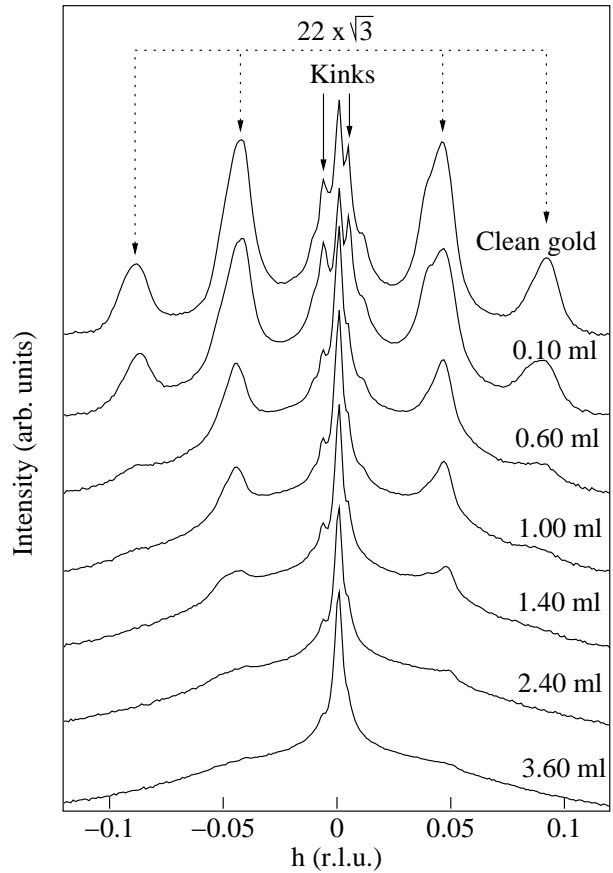}
\caption{\label{fig:RRAu111Scan}In-surface-plane HK scan around
the (0 1 L) rod at L=0.12 for clean gold (top) and for several Fe
deposits (the features of the $22\times\sqrt3$ reconstruction and
 those of the kinks are indicated).
squares.}
\end{figure}

Figure \ref{fig:RRAu111Map} shows a map around the (H K L) = (0 1
0.12) region of the reciprocal space at constant L of the clean
Au(111) surface. One can clearly see up to two orders of the
$22\times\sqrt{3}$ reconstruction. The signature of the
rectangular lattice produced by the kinks cannot be seen on this
representation but it is clearly visible in the top HK scan of
figure \ref{fig:RRAu111Scan} (this scan is referenced as (i) in
fig.\ref{fig:RRAu111}b) there are several faint peaks close to
the (0 1) reflection and one may also guess some faint kinks in
the first order reconstruction peaks, which actually correspond to
the projections of the rectangular kink lattice reflections on the
scanned axis.

\begin{figure}[htbp]
\includegraphics[width=8cm,bb=35 260 560 590]{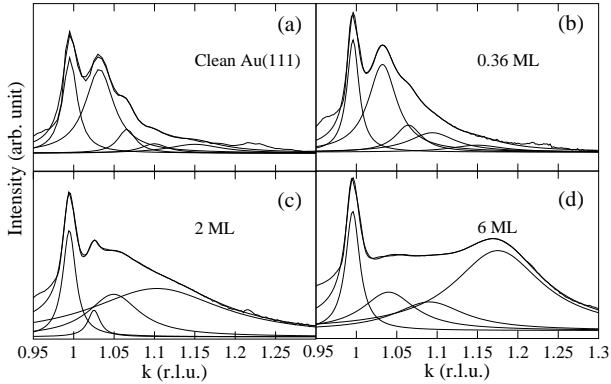}
\caption{\label{fig:RRAu111ScanK}K-scan along the (K, 0, 0.12)
direction with respect to the Fe thickness. (a) for the clean Au(111) surface. The first peak at K=1 corresponds to
the intersection of the crystal truncation rod with the surface plane, the second and
following are the different orders of the reconstruction. (b) for
0.36 ML Fe, (c) for 2.0 ML and (d) for 6.0 ML. \newline Note that the
small bump at K$\approx$1.225 is always present on clean gold and
likely belongs to a residual disoriented crystallite.}
\end{figure}

In figure \ref{fig:RRAu111ScanK}a, representing a K-scan at H=0
and L=0.12 (referenced as (ii) in fig.\ref{fig:RRAu111}b) for
clean gold, one can identify several features: the Au
crystal truncation rod at 2.88 {\AA$^{-1}$} (K=1) momentum
transfer, and the projection on the [H=0] direction of the first,
second and third orders of the $22\times \sqrt3$ reconstruction.
Due to the strong interatomic distance inhomogeneity, the
contributions are rather broad, and it is difficult to
quantitatively determine the modifications of the X-ray
diffraction scans upon Fe deposition at very low coverage. The K-scans
were therefore fitted by different lorentzian contributions.
For clean gold, the K-scan was simulated by the superposition of 5
lorentzian peaks (figure \ref{fig:RRAu111ScanK}a), one
for the Au crystal truncation rod, three for the
first, second and third order peaks  of the $22\times \sqrt3$ reconstruction.
The fifth peak was added to take into account the fourth and fifth order, and the background.
The presence and positions of the different orders of the gold reconstruction peaks were 
unambiguously determined from the HK-scan of figure \ref{fig:RRAu111Scan}.
The contribution  due to the kinks was not considered in the K-scan fits because they give only faint
projections.

\section{Fe deposition}

It is well known that the kink positions of the herringbone
reconstruction act as preferential nucleation sites for most $3d$
metals.\cite{Voigtlaender,Stroscio,Chambliss,Meyer,Fonin} In the
case of Fe, this leads to the growth of monolayer-high clusters,
located at the kinks, expanding laterally with increasing coverage.
\cite{Stroscio}

Upon Fe deposition, the main effect is the reduction of the
intensity of the reconstruction peaks (figure \ref{fig:RRAu111Scan}).
At about 0.5 ML, the second order peak has nearly vanished. The
first order reconstruction peak can however be observed up to about 2 ML.
This means that the reconstruction is progressively lifted, but
the surface is still over close-packed since the first order peak
remains. Interestingly, the features due to the rectangular kink
lattice remain present. Intuitively one would argue that these
features can only be present if there is a $22\times \sqrt3$
reconstruction. Since they do not disappear together with the
reconstruction, one must assume that the Fe clusters nucleated at
the kinks somehow help the Au substrate to keep the memory of the
rectangular lattice. A possible mechanism would be the
introduction of a periodic strain field in the gold substrate
during the growth. A similar mechanism has indeed been proved
recently in the case of N/Cu(100) \cite{Croset02} and was also
suggested by Grazing Incidence Small Angle X-ray Scattering
experiments performed on self-organized Co/Au(111) clusters.\cite{Fruchart03}

\begin{figure}[htbp]
\includegraphics[width=8cm,bb=30 100 560 730]{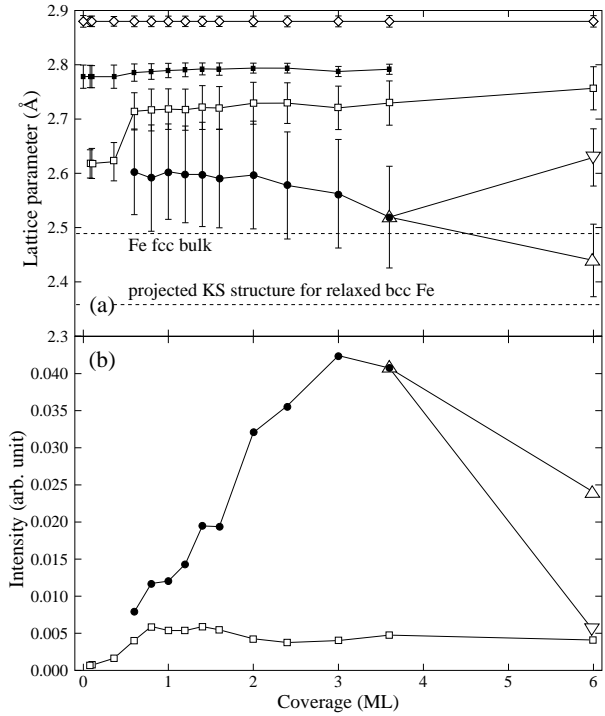}
\caption{\label{fig:Distances}(a) Real space distances derived from
the different contributions deduced from the fits of the K-scans
(fig. \ref{fig:RRAu111ScanK}) as a function of the Fe coverage. The
widths of the different contributions are plotted in form of
"error bars" (the actual error bars are $\sim5\%$). (b) Intensity of the corresponding fcc and bcc Fe contributions. ($\diamond$) represent the bulk gold, ($\blacksquare$) the gold surface reconstruction, ($\square$) the pseudomorphic fcc Fe, ($\bullet$) the relaxed fcc Fe. The open triangles stand for the bcc Fe (see text for details). The positions of the bulk fcc and the projected bcc iron are indicated by dashed lines. }
\end{figure}

The positions, widths and intensities of the CTR and the first
order reconstruction peaks are easily determined through the fit
procedure of the K-scans. Importantly, the rocking scans show that
the positions, even for higher reconstruction orders, do not
change at all with increasing coverage. Up to 2 ML the K-scans are
dominated by the first and second contributions that do not vary
in position. Conversely, the three other contributions show
intensity and position changes (figure \ref{fig:RRAu111ScanK}).
Since the second and next orders of the reconstruction peaks
quickly disappear in the type (i) HK-scans (figure
\ref{fig:RRAu111Scan}), we can confidently attribute the increase
of the three additional contributions to the Fe clusters. There
are only minor changes below 1 ML, indicating that the Fe atoms
grow in near registry with the underlying gold atoms. From the
fits, one can locate more precisely the different contributions.
In figure \ref{fig:Distances}a, the corresponding real space
distances of the different contributions are plotted as a function
of the Fe coverage. The widths of the different contributions are
plotted as "error bars". This graph confirms that the CTR and the
first order reconstruction peaks remain nearly unchanged in
lattice parameter, one corresponding to the bulk parameter (2.88
{\AA}), the other to the average nearest-neighbor distances
between gold atoms at the surface (2.78 {\AA}). The third
contribution ($\square$) shows a slight and sudden increase of the
lattice parameter at about 0.4 ML, from about 2.62 {\AA} to $\sim
$2.72 {\AA} before staying more or less constant with increasing
thickness. Between 0.4 ML and 3 ML there is a fcc contribution
($\bullet$) with a slightly decreasing parameter
and a very large width in K.

\begin{figure}[htbp]
\includegraphics[width=7cm,angle=0,bb=160 80 420 810]{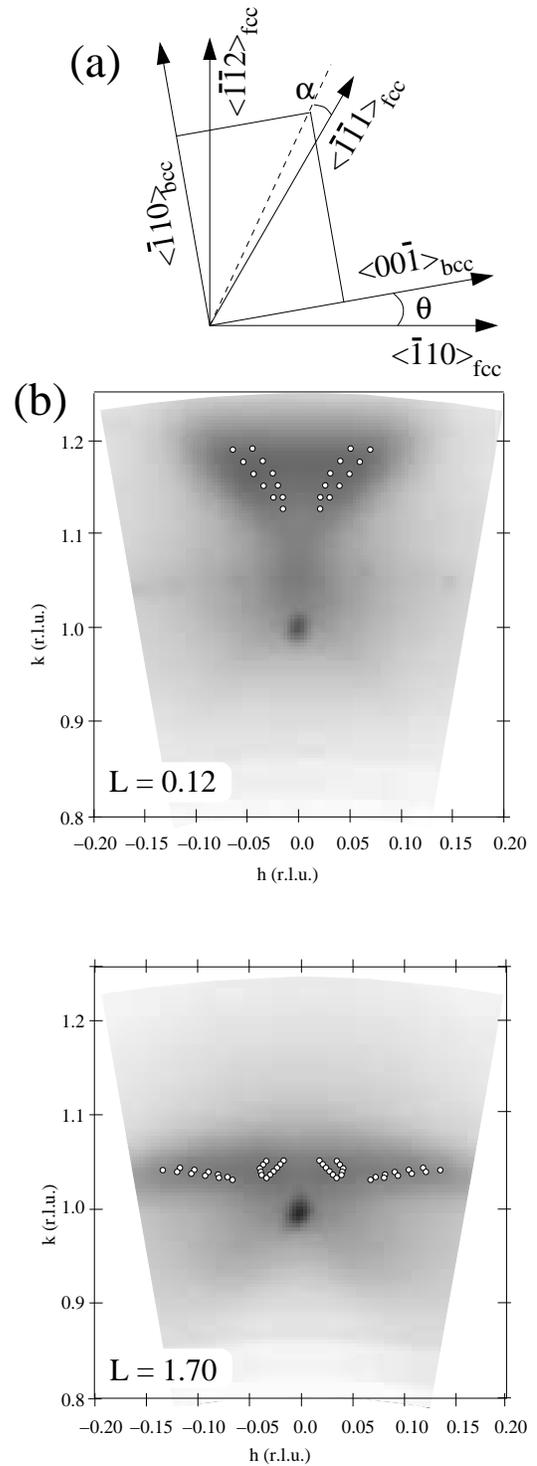}
\caption{\label{fig:Maps}
(a) bcc(110) lattice orientation with respect to the fcc (111) substrate
for the Nishiyama-Wassermann ($\theta=0$) and Kurdjumov-Sachs ($\alpha=0$)
orientations (the spots are originating from respectively three and six types of
symmetry equivalent domains). 
(b) Maps around the (H K L) = (0 1 0.12) and (H K L) = (0 1 1.7) regions of
the reciprocal space for 6.0 ML Fe. Simulated reflections (white dots) for a spread of
parameters and $\theta$ angles are superimposed. They give an indication of the rotational and parametric disorder.}  
 
\end{figure}

Above 3 ML, the fcc peak ($\bullet$) splits progressively into two large contributions ($\bigtriangleup$,$\bigtriangledown$), 
which correspond to a phase transition towards the stable bulk Fe
bcc(110) phase. The epitaxial relationship corresponds to an intermediate orientation between the Kurdjumov-Sachs (KS)
and the Nishiyama-Wassermann (NW) orientations (figure \ref{fig:Maps}a).  Since the global symmetry corresponds to the
KS symmetry, we will refer to this particular orientation as KS.
Diffracted intensity maps around (0 1 0.12) and (0 1 1.7) at constant L
show indeed typical patterns, although fuzzy, for such an
orientation (figure \ref{fig:Maps}b) (see ref. \onlinecite{Boukari99} and refs. therein). 
Simulations of the peak positions in the reciprocal space reproduce the experimental data rather well (see discusion). This structure induces 
a projected contribution in the K-scan at about (0.05 1.17 0.12).

\section{Discussion}

From the results above, one immediatly notices that the crystalline structure of the Fe
clusters evolves as a function of coverage. The individual contributions in the diffraction scans deserve
to be discussed separately.

The first contribution (($\square$) in figure
\ref{fig:Distances}b) exhibits a linear intensity increase with
the Fe thickness and then becomes constant after $\sim$1 ML. Its average
lattice parameter varies from $\sim$2.62
 \AA\ to 2.72 \AA\  and then remains almost constant. 
This behavior results from the competition between the Fe-Fe and Au-Fe interactions.
The Fe-Fe interactions tend to favor small Fe-Fe interatomic distances. 
Moreover, in the case of clusters, the edge atoms emphasize this effect\cite{step}.
Conversely, the Au-Fe interactions tend to impose an iron registry growth.
In the 0--0.4 ML thickness range, the nucleation takes place on the kinks, where the interatomic 
distances are the smallest (2.60 to 2.65 {\AA}\cite{BGO02}).
While growing laterally upon increasing the coverage, they spread over
the hcp or fcc regions where the gold atoms are separated by 2.80
to 2.85 {\AA}, what leads to an increased mismatch.
The 1D-coalescence ($\sim$ 0.4 ML) of adjacent islands along $<\bar{1}\bar{1}2>$ directions produces a sudden loss of edge atoms.
This latter effect, associated with the large mismatch between Fe and Au  at this coverage, enhances the weight of Fe-Au interactions leading to the observed transition.

A second Fe phase (($\bullet$) in figure \ref{fig:Distances}b)
with a slightly smaller lattice parameter starts growing at about
0.5 ML up to 3 ML. The intensity increase is proportional to the
additional Fe thickness showing that all incoming Fe atoms adopt
this crystallographic configuration. This is concomitant to 
the onset of the growth of a second Fe layer, with a slightly
relaxed parameter on top of the first epitaxial Fe layer: indeed,
the onset of the second layer growth starts between 0.5 and 0.8 ML
as noticed by STM experiments \cite{Stroscio}. In this thickness range (0.5-3 ML) the Fe film
must be understood as a relaxed Fe fcc layer on an interfacial
pseudomorphic layer.

Above 3-4 ML, the strain in the Fe layer starts to relax through a progressive
transformation into the bcc Fe phase with the KS bcc(110)//fcc(111)
epitaxial relationship. Indeed, this epitaxy does not yield any
Bragg position along our K-scans (figure \ref{fig:RRAu111ScanK})
and the contribution observed at K$\approx$1.17 is only due to the
tail of the bcc Bragg peaks located outside the K-scan. This
interpretation fairly well explains the decrease, and finally
splitting into two distinct bcc ($\bigtriangleup$,$\bigtriangledown$) contributions of the
fcc ($\bullet$) signal in figure \ref{fig:Distances}b. 
The ($\bigtriangleup$) contribution corresponds to the projection  of the KS structures, seen on the figure \ref{fig:Maps}b, on the $\vec{K}$ direction. 
The ($\bigtriangledown$) feature results from the crossing of the two KS structures on the $\vec{K}$ axis.
At large coverage the Fe film consists of a bcc layer on top of an
interfacial pseudomorphic 1 ML thick fcc layer. It is interesting to note that the fcc to bcc transition
occurs once the fcc lattice parameter has reached the value adopted by the bulk $\gamma$-Fe phase.
The somewhat blurred diffraction patterns in figure \ref{fig:Maps}b 
are due to both rotational disorder in the $\theta$ angle (as defined in \ref{fig:Maps}a)
and to a parametric spreading in the bcc $<\bar{1}10>$ direction,
from about 2.50 \AA\ to 2.65 \AA. For the highly strained domains, the orientation is
rather NW ($\theta \approx\ 0$), with nearly no orientational disorder, whereas
for the less strained domains, $\theta$ corresponds roughly to half of the ideal angle 
expected for the KS orientation, with an angular distribution of about 1$^\circ$.   
This can be seen in figure \ref{fig:Maps} where the reflection for different orientations
and parameters are superimposed to the experimental maps.

\section{Conclusion}

The structure and growth of Fe deposits on reconstructed Au(111) have been
investigated by GIXD. Up to one monolayer, Fe grows in close registry with the
reconstructed Au(111) substrate. However, due to the great inhomogeneity of the gold interatomic
distances at the surface, slight changes in the in-plane atomic distances already occur during
the lateral expansion of the monolayer-high Fe clusters upon growth. We qualify the structure as 
"local pseudomorphism". At the coalescence of the clusters along $<\bar{1}\bar{1}2>$, there
is a slight expansion of the average lattice parameter of the Fe clusters.
The second Fe layer grows, with a slightly reduced parameter.
Above 3 ML the film transits progressively and breaks into partly relaxed bcc(110) domains, in an orientation
intermediate between the Kurdjumov-Sachs and Nishiyama-Wassermann epitaxial relationships.
These structural results are consistent with the XMCD observations
\cite{Ohresser01} where strong changes in the magnetic properties
were evidenced. In particular, the magnetic spin moment showed a
sharp increase at the unidimensional coalescence around 0.4 ML
(going from $\sim$1.4$\mu_{B}$ to $\sim$2.4$\mu_{B}$). This change
can now be correlated to the increase of the lattice parameter.
Indeed, it is well known that for fcc Fe magnetism and structure
are linked with, in particular, a tendency to high magnetic spin
values for larger atomic volume. Moreover to fit the evolution of
the magnetic anisotropy energy with the coverage a N\'{e}el model
was used in reference \onlinecite{Ohresser01} assuming a
pseudomorphic growth of the first Fe layer. This assumption is now
further experimentally confirmed by the present GIXD results.

\begin{acknowledgments}
The authors are pleased to acknowledge the ID03 beamline staff of
the ESRF for excellent experimental conditions and support during
these experiments. We would also like to thank O. Fruchart for fruitfull discussion and A. Tagliaferri for his technical help.
\end{acknowledgments}

\end{document}